# Trapping of He Clusters by Inert-Gas Impurities in Tungsten: First-Principles Predictions and Experimental Validation

Duc Nguyen-Manh[1], S.L. Dudarev

CCFE, Culham Science Centre, Abingdon, Oxon, OX14 3DB, United Kingdom

**Abstract**

Properties of point defects resulting from the incorporation of inert-gas atoms in bcc tungsten are investigated systematically using first-principles density functional theory (DFT) calculations. The most stable configuration for the interstitial neon, argon, krypton and xenon atoms is the tetrahedral site, similarly to what was found earlier for helium in W. The calculated formation energies for single inert-gas atoms at interstitial sites as well as at substitutional sites are much larger for Ne, Ar, Kr and Xe than for He. While the variation of the energy of insertion of inert-gas defects into interstitial configurations can be explained by a strong effect of their large atomic size, the trend exhibited by their substitutional energies is more likely related to the covalent interaction between the noble gas impurity atoms and the tungsten atoms. There is a remarkable variation exhibited by the energy of interaction between inert-gas impurities and vacancies, where a pronounced size effect is observed when going from He to Ne, Ar, Kr, Xe. The origin of this trend is explained by electronic structure calculations showing that *p*-orbitals play an important part in the formation of chemical bonds between a vacancy and an atom of any of the four inert-gas elements in comparison with helium, where the latter contains only $1s^2$ electrons in the outer shell. The binding energies of a helium atom trapped by five different defects (He-v, Ne-v, Ar-v, Kr-v, Xe-v, where v denotes a vacancy in bcc-W) are all in excellent agreement with experimental data derived from thermal desorption spectroscopy. Attachment of He clusters to inert gas impurity atom traps in tungsten is analysed as a function of the number of successive trapping helium atoms. Variation of the Young modulus due to inert-gas impurities is analysed on the basis of data derived from DFT calculations.

## 1. Introduction

Generation of helium (He) in materials through transmutation nuclear reactions under neutron irradiation, giving rise to radiation swelling and grain boundary embrittlement, is a major factor limiting the lifetime of structural materials in fusion and fission power plants [1]. So far, experimental and theoretical effort has been focused primarily on the synergetic effects associated with the simultaneous accumulation of helium and hydrogen. Effects associated with the incorporation of other inert gases have not received much attention even though the agglomeration of noble gas atoms in metals and alloys is a well-known phenomenon observed in multi-beam ion implantation experiments [2]. One of the new ideas is to consider helium co-implantation with other inert-gas ions to simultaneously model dpa damage and dissolved noble gases, and through that to investigate the effect of neutron irradiation on nuclear fusion reactor components. However, a question arises if other noble gas atoms, such as Ar or Xe, are similar to He in relation to ion implantations. Reliable experimental data on defect energies for these noble gases in fusion structural materials, in particular their binding energies with vacancies or vacancy clusters, are very scarce. The effect of noble gas incorporation in these materials has not been systematically studied even though the agglomeration of noble gas atoms implanted in metals and alloys is often observed by transmission electron microscopy. Traditional view is that any noble gas impurity should be similar to helium, since there is no chemical interaction between an inert gas atom and a solid, and hence the behaviour of noble gas atoms in a material can potentially be explained assuming that local lattice distortion increases monotonically from He to Xe. Therefore, the development of an accurate predictive model based on first-principles calculations for defects formed due to the accumulation of helium and other noble gases in the crystal lattice of iron, steel, a non-magnetic body-centred cubic (bcc) refractory metal or alloy, is an issue of significance for the quantification of radiation damage effects on structural integrity of fusion reactor components. In this paper, we investigate small helium clusters trapped by inert-gas impurities in tungsten, and validate our theoretical study by comparison with experimental thermal desorption spectroscopy data.

---

[1] Corresponding author. Tel.: +44-1235-466-284; fax: +44-1235-466-435.
*E-mail address*: duc.nguyen@ccfe.ac.uk.

## 2. Benchmarking density functional theory (DFT) calculations

First-principles calculations described in this work were performed using density functional theory (DFT) implemented in the Vienna Ab-initio Simulation Package (VASP) code with the interaction between ions and electrons described by the projector-augmented waves (PAW) method [4]. The Perdew–Burke–Ernzerhof (PBE) generalized gradient approximation (GGA) [5] is used for describing the effects of exchange and correlation in tungsten as well as in all the inert-gas elements through the application of standard PAW potentials. For diatomic noble gas molecules, the PBE model predicts interaction energies and equilibrium distances [6-7] closer to both the theoretical hybrid functional PBE0 and experimental values in comparison with the Perdew-Wang (PW) functional [8] used in previous studies of He in bcc-W [9-10]. It is important to emphasize that the inclusion of *semi-core electrons* in the valence states has a significant effect on the energetic trend of intrinsic radiation-induced self-interstitial atom (SIA) defects in all the bcc transition metals [11-12]. In particular, we find that in all the non-magnetic bcc transition metals the most stable <111> configuration of SIAs cannot be reproduced correctly if the semi-core electrons are ignored, this is especially significant for transition metals of group 6B (Cr, Mo, W). All the total energy calculations were performed using a $4 \times 4 \times 4$ (128 atom) bcc super-cell, with plane-wave cut-off energy of 400 eV and *k*-point mesh with spacing of 0.15 (Å)$^{-1}$. In elastic constant calculations, the *k*-point mesh spacing was 0.10 (Å)$^{-1}$.

Table 1. DFT formation and binding energies (in eV) of intrinsic and He defects in bcc-W computed using the PBE exchange-correlation functional (present work), compared with those calculated using the PW functional from Ref. [9-10]. The PBE-predicted relaxation volumes of defects, given in the units of tungsten atomic volume ($V_0$) in a perfect bcc 4x4x4 bcc super-cell, are given in the last column.

| Defect-type | PBE (Present work) | PW ([9-10]) | Relaxation volume ($V_0$) |
|---|---|---|---|
| $E_f$(vac) | 3.24 | 3.19 | -0.37 |
| $E_f$(SIA<111>) | 10.06 | 10.53 | 0.68 |
| $E_f$(He_sub) | 4.83 | 4.70 | -0.23 |
| $E_f$(He_tet) | 6.22 | 6.16 | 0.36 |
| $E_f$(He_oct) | 6.44 | 6.38 | 0.37 |
| $E_b$(2He) | 1.03 | 1.03 | 0.80 |
| $E_b$(3He) | 1.18 | 1.36 | 1.16 |
| $E_b$(4He) | 2.23 | 1.52 | 1.65 |
| $E_b$(5He) | 1.59 | 1.64 | 2.03 |
| $E_b$(He-v) | 4.63 | 4.57 | -0.23 |
| $E_b$(2He-v) | 3.10 | 3.11 | -0.06 |
| $E_b$(3He-v) | 3.33 | 3.28 | 0.14 |
| $E_b$(4He-v) | 3.26 | 2.61 | 0.38 |
| $E_b$(5He-v) | 1.88 | 1.44 | 0.71 |
| $E_b$(6He-v) | 2.18 | 2.08 | 1.10 |

For benchmarking our DFT-PBE calculations, we have cross-checked the formation energies of intrinsic defects and substitutional and interstitial helium defect configurations, as well as binding energies of He clusters and their interaction with a mono-vacancy in bcc-W, as shown in Table 1. The results are compared with the corresponding data obtained from DFT calculations for the same 128-atom super-cell but using the PW exchange-correlation functional [9-10]. We find significant differences between the predicted binding energies of 4-He cluster (from $He_{int}$+3He reaction) : 2.23 eV versus 1.52 eV for PBE versus PW and also for 4-He-atom cluster interacting with a mono-vacancy (from $He_{int}$+3H-v reaction): 3.24 eV and 2.61 eV for PBE and PW, respectively. It is interesting that in the assessment of helium cluster behaviour in tungsten [13], the binding energy of 4He-v defect complex was altered from 2.61 eV to the value close to 3.0 eV, which is in better agreement with our DFT-PBE value of 3.24 eV. The relaxation volumes of defects, i.e. volume differences between the fully relaxed defect configurations and perfect lattice structures, are given in the last column of Table 1. The calculated relaxation volumes for vacancies and interstitial helium defects are in good agreement with literature data [14-15]. Table 1 shows that the relaxation volume of nHe-v defects changes sign from negative to positive as the number of He atoms, occupying a vacancy, increases. DFT data on relaxation volumes have been used for interpreting lattice expansion of tungsten alloys implanted with He at low temperature.

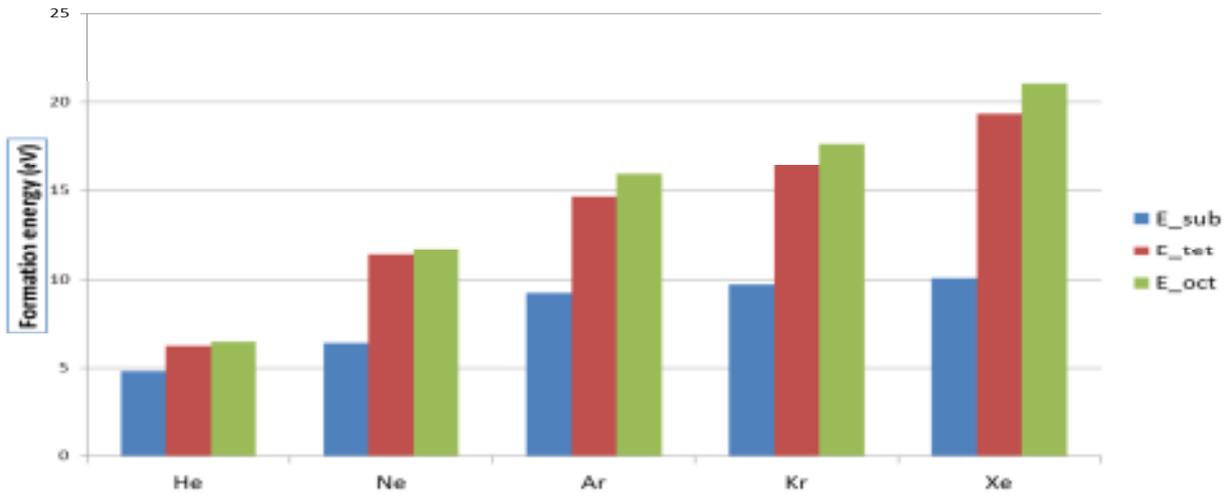

Fig. 1. Formation energies of inert-gas atoms (He, Ne, Ar, Kr, Xe) for substitutional, tetrahedral and octahedral configurations in bcc-W.

Formation energies of He in substitutional and interstitial positions in tungsten found in our simulations are in good agreement with literature data. Fig. 1 shows a trend exhibited by these formations energies when going from He to Ne, Ar, Kr and Xe. Our calculations show that the most stable interstitial configuration for the latter four inert-gas elements in bcc-W is the tetrahedral site. The formation energy difference between octahedral and tetrahedral configurations increases from 0.22 eV for He to 0.51, 1.13, 1.25, and 1.62 eV for Ne, Ar, Kr and Xe, respectively, indicating the dominant contribution from lattice distortion to the defect formation energy as the size of an inert gas atom increases from helium to xenon. The octahedral site is a metastable configuration with respect to the tetrahedral one and therefore the energy difference between the two configurations provides an estimate for the migration barrier of interstitial inert-gas atoms in bcc-W. Energy difference between the two interstitial configurations for inert-gas atoms in tungsten is quite sensitive to size of super-cell used for DFT calculations. A recent study [16] of three inert-gases (He, Ne and Ar) defects in bcc-W modelled using a small 54-atom supercell gives the interstitial energy difference of 0.38 eV for the helium case (in comparison with 0.22 eV found here and in previous studies [9-10]), whilst for neon and argon, this energy difference is 0.62 eV and 1.06 eV compared to 0.51 eV and 1.13 eV found in our 128-atom super-cell study, respectively. Our nudged elastic band calculations give the migration barrier of 0.07, 0.14 and 0.23 eV for the tetrahedral interstitial He, Ne and Ar in a comparison with 0.06, 0.17 and 0.19 eV, respectively, found in [14]. The calculated formation energies of inert-gas atoms in substitutional positions shown in Fig. 1 exhibit a similar monotonic energy trend from He to Xe. The substitutional formation energy is 4.83 eV (compared with 4.70 eV from [9-10]) for He and the corresponding predicted values for Ne, Ar, Kr and Xe are 6.25, 9.28, 9.74 and 10.09 eV, respectively. It is clear from Fig. 1 that these values are not of the same magnitude as the formation energies of inert-gas defects in either of the two interstitial configurations, especially for Ne, Ar, Kr and Xe. As discussed in Refs. [1,17-18], it is not sensible to compare directly the formation energies of substitutional configurations with those obtained for the interstitial sites. Strictly speaking, the definition of the solution energy involves a two-step process: first creating a vacancy and then inserting an inert-gas atom into it. In agreement with a previous study [16], we find, from calculations involving full relaxation of atomic positions, that a He defect prefers to occupy a substitutional position rather than an octahedral interstitial configuration away from a vacancy site. For other inert-gas defects, due to the larger distortion around a vacancy associated with the insertion of an inert gas atom into it, the difference between the energies of the two configurations is relatively small although it is still in favor of a substitutional configuration. Therefore, in the analysis of interaction between a vacancy and an inert-gas impurity we consider the substitutional configuration as the lowest-energy ground state of the system.

## 3. Trapping of helium at substitutional inert-gas impurities

Binding energy $E_b(A,B)$ between two entities A and B is defined as the total energy difference between the energies $E(A)$, $E(B)$ of supercells containing A and B separately, and those of a supercell containing both A and B, $E(A+B)$ and a reference energy of a super-cell without A and B. From this definition, positive binding energy indicates attraction between the entities. The calculated energy of binding between a mono-vacancy (v) and an inert-gas atom are 4.63, 8.18, 8.54, 9.98 and 12.53 eV for He-v, Ne-v, Ar-v, Kr-v and Xe-v configurations, respectively. For He case, the energy of dissociation via interstitial migration is $E_D=E_b(He-v)+E_{mig}(He)=4.63+0.07=4.7$ eV, and is in very good agreement with the experimental value of 4.6 eV characterising this process in tungsten [19]. We found that the binding energy increases dramatically from He-v to Ne-v, Ar-v, Kr-v, and Xe-v, indicating strong attachment of inert-gas atoms to vacancies. The above characteristic trend of energies exhibited by He and other inert-gases in tungsten can be understood by comparing their atomic radii with the covalent ones derived from the Periodic Table. While for helium the atomic and covalent radii are comparable (31 vs. 32 pm), the former is much smaller then the latter one for Ne (38 vs. 69 pm), Ar (71 vs. 97 pm), Kr (88 vs. 110 pm) and Xe (1.08 vs. 130 pm). From the electronic structure point of view, the difference can be explained by the fact that the closed shells of He only contain $2s$ electrons whereas in other inert-gas atoms additional $p$ electrons are present. If an inert-gas atom is in a substitutional site in bcc-W, covalent interactions due to hybridization between the $p$ orbitals of the inert-gas atom and the $5d$-orbitals of tungsten atoms plays an important role, increasing the strength of chemical bonding. Fig. 2 shows the calculated partial electronic density of states of inert-gas impurity atoms in bcc-W, decomposed into $s$ and $p$ contributions. For He, a localized impurity state can be seen at a very low energy (-12 eV) and its contribution to the valence states near the Fermi energy originates from the $s$ orbitals only. This is different from PDOS of other inert-gas atoms, where the localized impurity state energies decrease from Ne to Xe and, at the same time, additional contributions from $p$ orbitals to the valence states become more important.

Due to the relatively small energy difference (0.07 eV) between the tetrahedral and octahedral defect configurations, helium injected into tungsten undergoes rapid interstitial diffusion at room temperature, unless it encounters lattice defects. Helium itself does not produce observable damage, but it can be trapped by defects produced by prior ion bombardment, for example, by inert-gas atoms. Thermal desorption spectroscopy (TDS) experiments [20] show that binding of helium atoms to such defects is characterised by well defined discrete energies, the magnitude of which depend on whether a vacancy is occupied by a single He atom (He-v), or Ne, Ar, Kr, or Xe impurity atoms. The binding energy of helium trapping at defects of vacancy type is therefore an important parameter for validating theoretical calculations through comparison with experimental measurements of trapping energies of He atoms in tungsten in the presence of lattice damage. The most important and pronounced discrete binding energy is related to the attachment of one helium atom to an inert-gas impurity. The binding energies of He trapped by five different inert-atom impurity defects: He-v, Ne-v, Ar-v, Kr-v and Xe-v are shown in Fig. 3. The trapping energies predicted by DFT calculations are 3.10, 2.47, 1.93, 1.66 and 1.24 eV. They agree well with experimental data derived from thermal desorption spectroscopy [20] of 3.1, 2.3, 1.9, 1.5 and 1.2 eV, respectively. The latter values are derived from the peak temperatures of He desorption rate dependence, treated as functions of temperature, caused by trapping of helium atoms by He-v, Ne-v, Ar-v, Kr-v, Xe-v defects. From Fig. 3, we see that the binding energy decreases monotonically as a function of the atomic number from He to Xe, as it was proposed in [20], with DFT calculations highlighting the important role played by covalent contributions to binding energies for Ne, Ar, Kr and Xe in comparison with He.

## 4. Trapping of He clusters by inert-gas impurities

TDS results for helium in tungsten suggested that all the desorption spectra can be explained assuming several discrete binding states, which give rise to peaks at various temperatures in the helium desorption spectra, corresponding to cases where damage results from inert-gas ion implantation followed by the injection of helium ions [21]. The successive addition of the second, third, fourth and fifth helium atoms to a trap already occupied by a helium impurity can be investigated systematically by DFT calculations of binding energies.

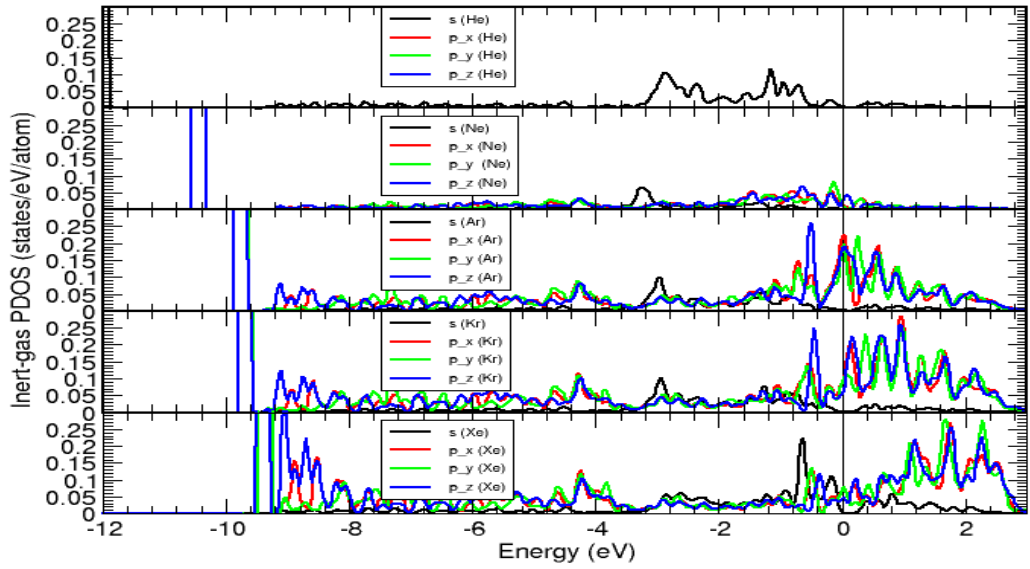

Fig.2. Partial electronic density of states for *s* and *p*-orbitals of inert-gas atom impurities calculated for substitutional configurations of inert-gas atoms in bcc-W. The Fermi level corresponds to the origin of the energy axis.

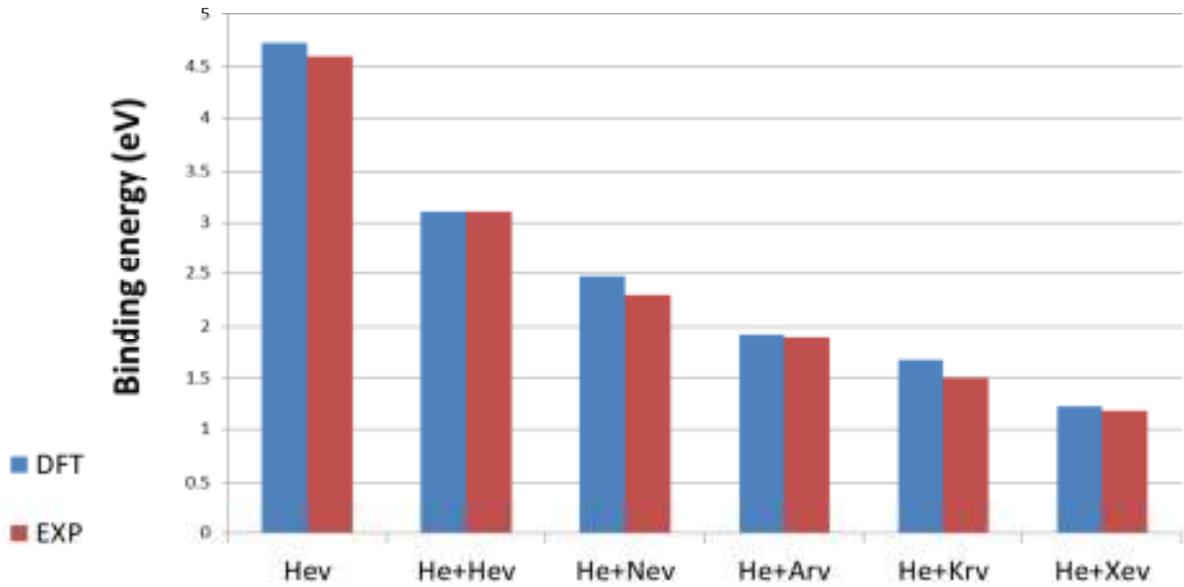

Fig. 3. Binding energies of He atom to a vacancy and to other substitutional inert-gas atoms in tungsten predicted by DFT calculations performed as a part of this study, compared to experimental data derived from thermal desorption spectroscopy [20].

Fig. 4 shows binding energies corresponding to multiple occupancy of various inert gas traps as a function of helium concentration. As Table 1 shows, the binding energy for reaction He+He$_{n-1}$-v -> He$_n$-v increases from 3.1 eV for the first atom trapped by a He-v impurity (2He-v), to 3. 33 (3He-v) and 3.26 (4He-v) eV for the second and the third helium atom, respectively. This trend shows that helium implanted in tungsten at room temperature forms a helium-vacancy cluster that predominately contains more than one helium atom per vacancy. This theoretical result is in agreement with the TDS data from Ref. [21] although the estimated experimental binding energy reported in [19] for 3He-v and 4He-v clusters is somewhat lower than the DFT value, namely, 2.9 and 2.4 eV, respectively. It is worth mentioning that apart from the first prounouced peak, the decomposition of helium desorption peaks into various helium cluster trapped configurations introduces some uncertainty in the evaluation of trapping energies. Our results for the fourth and fifth helium addition to a helium-vacancy cluster give the binding energies of 1.88 and 2.18 eV that are lower in a comparison with the second and the third one.

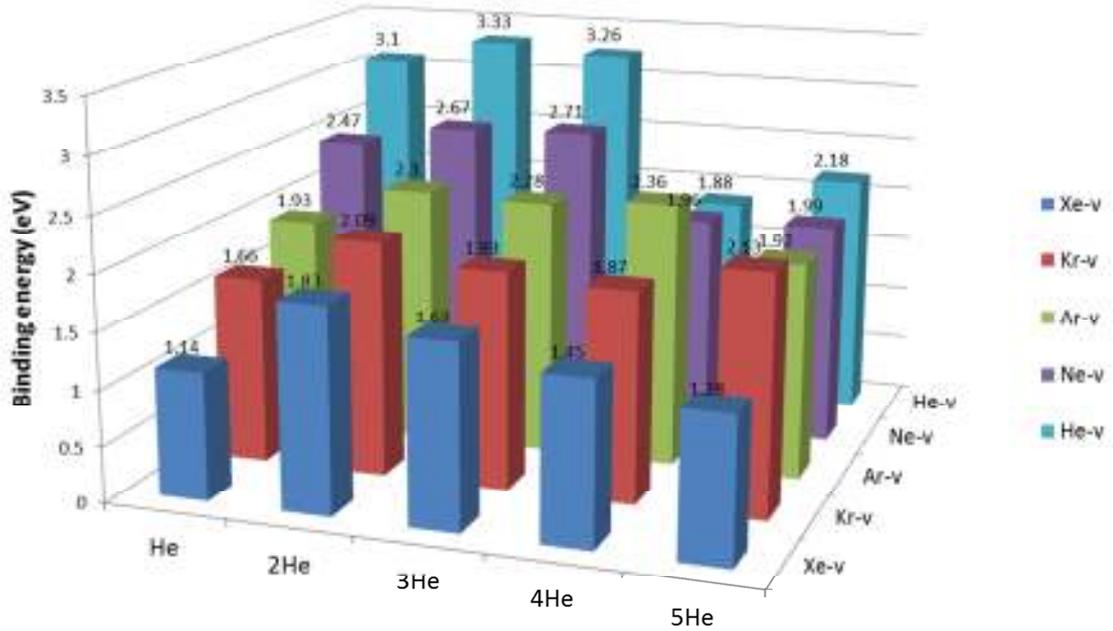

Fig. 4. Binding energies predicted by DFT-PBE calculations for a various number of helium atoms (up to 5 He) trapped by various inert-gas impurities in tungsten.

For Ne-v trap, the successive attachment of helium atoms gives rise to a similar trend as the He-v one, but the magnitude of binding energy decreases from 2.47 eV for the first helium atom to 1.99 eV for the fifth one. The change in the binding energy ($E_b$) as a function of helium atoms is however less significant for Ne-v trap than for He-v one, and this is more consistent with experimental information suggesting that $E_b$ tends to stay constant. For the Ar-v, Kr-v and Xe-v traps, a common trend found in DFT calculations is that $E_b$ tends to increase starting from the binding energy of the first helium atom attachment, and this trend is again consistent with experimental observations although the increase of $E_b$ is not quite monotonic as it is shown in Fig. 7 of [20]. Finally, for configurations with more than 10 helium atoms per trap, the TDS measurements suggest that the traps lose their identity and transform into helium bubbles. Recent molecular dynamic simulations performed using empirical potentials based on DFT for He clusters with different helium to vacancy ratios explain this experimental result [22].

## 5. The Young modulus changes due to inert-gas atom defects

Recently, there has been strong interest in understanding effects of helium and other ion irradiation on micro-mechanical properties of tungsten and tungsten alloys for fusion applications [23-25]. To understand the changes of elastic properties resulting from inert-gas impurity accumulation in tungsten, a fully relaxed 4x4x4 supercell was used to evaluate elastic constants for perfect lattice as well as for substitutional inert-gas defect configurations. Since inert-gas impurities are in substitutional sites, the cubic symmetry of the supercell in the presence of a defect is conserved and the three independent elastic constants can be evaluated using a standard methodology. A detailed description of the method used for calculating elastic constants in W and W alloys is given elsewhere [26-28]. The resulting bulk, shear, and the Young moduli as well as the Poisson ratio are given in Table 2 for the fully relaxed configurations.

Table 2. The bulk modulus, shear modulus, the Poisson ratio and the Young modulus predicted by DFT-PBE calculations of elastic stiffness matrices for inert-gas impurity substitutional defects together with reference values computed for a 4x4x4 W super-cell.

|         | Bulk modulus (GPa) | Shear modulus (GPa) | Poisson ratio | Young moulus (GPa) |
|---------|---------|---------|---------|---------|
| 4x4x4-W | 304.5 | 162.2 | 0.2738 | 413.2 |
| He-W    | 299.6 | 139.6 | 0.2984 | 362.5 |
| Ne-W    | 297.3 | 145.6 | 0.2894 | 375.6 |
| Ar-W    | 298.4 | 151.3 | 0.2832 | 388.2 |
| Kr-W    | 299.3 | 155.0 | 0.2794 | 396.6 |
| Xe-W    | 300.1 | 156.8 | 0.2775 | 400.6 |

Table 2 shows that all the elastic moduli characterising the inert-gas defects considered here decrease whereas the corresponding Poisson ratio increases slightly in comparison with pure tungsten. Further calculations of elastic moduli averaged over orientations of a single self-interstitial atom defect in a <111> configuration as well as of a 2He-v cluster in the same bcc-W supercell show that while the bulk modulus increases for the former case (312.9 GPa) and decreases for the latter (298.3 GPa), the Young modulus in both cases decreases from the bulk value (413.2GPa) to 376.2 GPA and 387.3 GPa, respectively, due to the increase of the Poisson ratio. Currently the change of the Young modulus in W and W alloys under inert-gas ion implantation is under investigation by various experimental methods in order to verify the above DFT predictions [29].

## 6. Conclusions

We have carried out a systematic study of inert-gas impurities and their interactions with small helium clusters in tungsten using DFT calculations in the GGA-PBE exchange-correlation approximation. Comparison of helium defect energies with previous DFT studies performed using the GGA-PW functional showed differences for the formation energy of a cluster of four interstitial He atoms in a tetrahedral configuration as well as for the binding energy of 4He-v cluster. The relaxation volume of nHe-v clusters changes sign from negative to positive for n>2. It is found that the most stable interstitial configuration of Ne, Ar, Kr and Xe is the tetrahedral site as in the case of He in tungsten. The binding energy between inert-gas atoms and a mono-vacancy in bcc-W increases strongly from He to other inert-gas atoms. This trend can be explained by the additional contribution coming from hybridization between $p$ orbitals of Ne, Ar, Kr and Xe with the $d$ orbitals of tungsten. The trapping energies of a helium atom attached to one of the six different impurity configurations: vacancy, He-v, Ne-v, Ar-v, Kr-v and Xe-v, predicted by DFT, are in excellent agreement with the binding energies evaluated from the most pronounced peaks of TDS spectra of helium desorption rates as functions of temperature. As successive helium atoms are trapped by the inert-gas impurities, the binding energies tend to increase for the second atoms for all the five inert-gas atom-vacancy traps. From the third to the fifth helium atom, multiple trapping tend to agree with experimental observations although the binding energy dependence as a function of He atom concentration is not monotonic for all the inert-gas traps. Finally, we have investigated the change in elastic properties of tungsten with and without inert-gas impurities and found that the Young modulus decreases due to the accumulation of inert-gas defects.

**Acknowledgements**

This work was part-funded by the RCUK Energy Programme (Grant Number EP/I501045) and by the European Unions Horizon 2020 research and innovation programme under grant agreement number 633053. To obtain further information on the data and models underlying this paper contact PublicationsManager@ccfe.ac.uk. The views and opinions expressed herein do not necessarily reflect those of the European Commission. This work was also part-funded by the United Kingdom Engineering and Physical Sciences Research Council via a programme grant EP/G050031. DNM would like to thank the International Fusion Energy Research Centre (IFERC) for providing access to a supercomputer (Helios) at Computational Simulation Centre (CSC) in Rokkasho (Japan). The authors would like to thank S.G. Roberts, F. Hofmann and A. De Backer for stimulating discussions.